\magnification \magstep1
\raggedbottom
\openup 2\jot
\voffset6truemm
\centerline {\bf ON THE BEHAVIOUR OF A RIGID CASIMIR CAVITY}
\centerline {\bf IN A GRAVITATIONAL FIELD}
\vskip 1cm
\noindent
Giuseppe Bimonte, Enrico Calloni, Luciano di Fiore,
Giampiero Esposito, Leopoldo Milano, Luigi Rosa
\vskip 1cm
\noindent
{\it Universit\`a di Napoli Federico II, Dipartimento
di Scienze Fisiche, Complesso
Universitario di Monte S. Angelo, Via Cintia, 80126 Napoli, Italy}
\vskip 0.3cm
\noindent
{\it Istituto Nazionale di Fisica Nucleare, Sezione di
Napoli, Complesso Universitario di Monte S. Angelo, Via Cintia,
80126 Napoli, Italy}
\vskip 1cm
\noindent
{\bf Abstract}.
A detailed investigation is presented of the energy-momentum
tensor approach to the evaluation of the force acting on a
rigid Casimir cavity in a weak gravitational field. Such a
force turns out to have
opposite direction with respect to the gravitational
acceleration. The order of magnitude for a multi-layer cavity
configuration is derived and experimental feasibility is
discussed, taking into account current technological resources.
\vskip 100cm
\leftline {\bf 1. Introduction}
\vskip 0.3cm
\noindent
Although much progress has been made in the evaluation and
experimental verification of effects produced by vacuum energy in
Minkowski space-time  it remains unclear why the observed universe
exhibits an energy density much smaller than the one resulting
from the application of quantum field theory and the equivalence
principle. Various hypotheses have  been put forward on the
interaction of virtual quanta with the gravitational field. For
instance some arguments seem to suggest that virtual photons do
not gravitate [1], while other authors have suggested
that Casimir energy contributes to gravitation [2]. So
far it seems fair enough to say that no experimental verification
that vacuum fluctuations can be treated according to the
equivalence principle has been obtained as yet, even though there
are expectations, as we agree, that this should be the case.

Motivated by all these considerations, our paper computes the
effect of a gravitational field on a rigid Casimir cavity,
evaluating the net force acting on it. The order of magnitude of
the resulting force, although not allowing an immediate
experimental verification, turns out to be compatible with the
current extremely sensitive force detectors, actually the
interferometric detectors of gravitational waves.  We evaluate the
force acting on this non-isolated system at rest in the
gravitational field of the earth by studying the regularized
energy-momentum tensor and recovering the same result as mode by
mode evaluation with the assumption that virtual photons do
gravitate and suffer gravitational read shift. The associated
force turns out to have opposite direction with respect to
gravitational acceleration. Orders of magnitude are discussed
bearing in mind the current technological resources as well as
experimental problems, and physical relevance of the analysis is
stressed.
\vskip 0.3cm
\leftline {\bf 2. Evaluation of the force}
\vskip 0.3cm
\noindent
In order to evaluate the force due to the gravitational field let
us suppose that the cavity has geometrical configuration of two
parallel plates of proper area $A = L^{2}$ separated by the proper
length $a$. The system of plates is taken to be orthogonal to the
direction of gravitational acceleration $\vec g$.

In classical general relativity, the force density can be
evaluated according to [3]
$$
f_{\nu}=-{1\over \sqrt{-{\rm det} \; g}} {\partial \over \partial
x^{\mu}} \Bigr(\sqrt{-{\rm det} \; g} \; T_{\; \nu}^{\mu}\Bigr)
+{1\over 2}{\partial g_{\rho \sigma}\over \partial x^{\nu}}
T^{\rho \sigma},
\eqno (1)
$$
where $T^{\mu \nu}$ is the energy-momentum tensor of matter,
representing the energy densities of all non-gravitational fields,
and gravity couples to $T^{\mu \nu}$ via the Einstein equations:
$R_{\mu \nu}-{1\over 2}g_{\mu \nu}R=8\pi G T_{\mu \nu}$. However,
we are eventually interested in quantum field theory in curved
space-time, where gravitation is still treated classically, while
matter fields are quantized. At this stage the key assumption is
that one can take the expectation value of the quantum
energy-momentum tensor in some state and evaluate the external
force as if it were a classical force. This should therefore
represent the expectation value of the ``quantum force'' in the
given state. According to this theoretical model, we have to find
the regularized energy-momentum tensor $\langle T^{\mu \nu}
\rangle$ and insert it into the right-hand side of Eq. (1) to find
eventually $\langle f_{\nu} \rangle$. If we were in Minkowski
space-time with metric $\eta$, we might exploit the result of Ref.
[4], according to which
$$
\langle T^{\mu \nu} \rangle ={\pi^{2}{\hbar}c \over 180 a^{4}}
\left({1\over 4}\eta^{\mu \nu}-{\hat z}^{\mu}{\hat z}^{\nu}
\right),
\eqno (2)
$$
where ${\hat z}^{\mu}=(0,0,0,1)$ is the unit spacelike 4-vector
orthogonal to the plates' surface. If such an analysis is
performed for an accelerated but locally orthonormal reference
frame in curved space-time, corresponding to our system at rest in
the earth's gravitational field, the Minkowski line element should
be replaced by the line element near the observer's world line and
this reads (neglecting possible rotation effects)
$$
ds^{2}=-(1+2A_{j}x^{j})(dx^{0})^{2}+\delta_{jk}dx^{j}dx^{k}
+O_{\alpha \beta}(|x^{j}|^{2})dx^{\alpha}dx^{\beta}.
\eqno (3)
$$
Here $c^{2}{\vec A}$ (with components $(0,0,|{\vec g}|)$), the
observer's acceleration with respect to the local freely falling
frame, shows up in the correction term $-2A_{j}x^{j}$ to $g_{00}$,
which is proportional to distance along the acceleration
direction. Note that first-order corrections to the line element
are unaffected by space-time curvature. Only at second order,
which is beyond our aims, does curvature begin to show up. With
this understanding we find, on setting
$$
K \equiv {\pi^{2}{\hbar}c \over 180 a^{4}},
\eqno (4)
$$
that the non-vanishing components of the regularized
energy-momentum tensor which contribute to $\langle f_{z} \rangle$
are
$$
\langle T_{\; 3}^{3} \rangle = \langle T^{33} \rangle =-{3\over
4}K ; {}   \langle T^{00} \rangle =-{K\over 4}{1\over 1+2Az}.
\eqno (5)
$$
Since (of course $x^{3} \equiv z$ and hence $\langle f_{3}
\rangle$ denotes $\langle f_{z} \rangle$)
$$
\langle f_{z} \rangle = -{1\over \sqrt{1+2Az}}{\partial \over
\partial x^{\mu}}\Bigr(\sqrt{1+2Az}T_{\; 3}^{\mu}\Bigr)
+{1\over 2}{\partial g_{\eta \rho}\over \partial z} T^{\eta \rho}
=\left({3\over 4}K+{1\over 4}K \right){A\over 1+2Az},
\eqno (6)
$$
we find for the component of the full force along the $z$-axis the
formula
$$
\langle F_{z} \rangle \cong aL^{2} \langle f_{z} \rangle \cong
aL^{2}gK,
\eqno (7)
$$
where integration over the volume $V=aL^{2}$ has reduced to a
simple multiplication by $aL^{2}$ to our order of approximation.

Interestingly, the resulting force has opposite direction with
respect to the gravitational acceleration. Note also that the
force density in Eq. (6), and hence the force itself in Eq. (7),
is the sum of two terms: the latter, proportional to $g{K\over
4}$, arises from the Casimir energy encoded into $T^{00}$, and can
be interpreted as the Newtonian repulsive force (i.e. ``push'') on
an object with negative energy. The former term, proportional to
${3\over 4}gK$, results from the pressure along the acceleration
axis and can be interpreted as the mass contribution of the
spatial part of the energy-momentum tensor. With this
understanding, our result agrees with the equivalence principle,
according to which for every pointlike event of space-time, there
exists a sufficiently small neighbourhood such that in every
local, freely falling frame in that neighbourhood, all laws of
physics obey the laws of special relativity (as a corollary, the
gravitational binding energy of a body contributes equally to the
inertial mass and to the passive gravitational mass). In
particular, we agree with the statement that a system with a given
rest energy momentum tensor $T^{{\overline {00}}}$ has the
inertial mass tensor $m^{ij}=T^{{\overline {00}}}\delta^{ij}
+T^{{\overline {ij}}}$.

Furthermore, it is very important
to note that, when considering the total force acting on the real
cavity, which is an isolated system, the contribution to the force
resulting from the spatial part of the energy-momentum tensor is
balanced by the contribution of the mechanical energy-momentum
tensor, and hence should not be considered for experimental
evaluation. The resulting force is then the Newtonian force on the
sum of the rest Casimir energy and rest mechanical mass: the
contribution of vacuum fluctuations leads to a gravitational push
on the Casimir apparatus expressed by the formula
$$
{\vec F} \cong {\pi^{2}L^{2}{\hbar}c \over 720 a^{3}} {g\over
c^{2}}{\vec e}_{r},
\eqno (8)
$$
which should be tested against observation. As far as we can see,
our calculation suggests that the electromagnetic vacuum state in
a weak gravitational field is red-shifted, possibly adding
evidence in favour of virtual quanta being able to gravitate, a
non-trivial property on which no universal agreement has been
reached in the literature [1,2], as we
already noticed in the introduction. To further appreciate this
point, we now find it appropriate and helpful to show that the
same result can be derived from a mode-by-mode analysis. Recall
indeed that in Minkowski space-time the zero-point energy of the
system can be evaluated, in the case of perfect conductors, as
$$
U={\hbar c L^{2}\over 2}\sum_{n=-\infty}^{\infty} \int {d^{2}k
\over (2\pi)^{2}}\sqrt{k^{2}+ \left({n \pi \over a}\right)^{2}},
\eqno (9)
$$
where we have considered the normal modes labelled by the integer
$n$ and the transverse momentum $k$. On performing the integral by
dimensional regularization, the energy takes the well known
Casimir expression
$$
U_{\rm reg}=-{\pi^{2}L^{2}{\hbar}c \over 720 a^{3}},
\eqno (10)
$$
where the final result is independent of the particular
regularization method.

Consider next the cavity at rest in a Schwarzschild geometry. On
assuming that virtual quanta do gravitate, and bearing in mind
that the modes remain unchanged, the energy of each mode is
red-shifted by the factor $\sqrt{g_{00}}=\sqrt{1-{\alpha \over
r}}$, with $\alpha \equiv {2GM\over c^{2}}$. Hence the total
energy can be written as (the suffix $S$ referring to
Schwarzschild)
$$
U_{S}={\hbar c L^{2}\over 2}\sum_{n=-\infty}^{\infty} \int {d^{2}k
\over (2\pi)^{2}}\sqrt{\left(k^{2}+ \left({n \pi \over
a}\right)^{2}\right)(\sqrt{g_{00}})^{2}}.
\eqno (11)
$$
On performing dimensional regularization one finds therefore
$$
(U_{S})_{\rm reg}=(g_{00})^{1\over 2}U_{\rm reg} =-\left(1-{\alpha
\over r}\right)^{1\over 2} {\pi^{2}L^{2}{\hbar}c \over 720 a^{3}}.
\eqno (12)
$$
Now we assume that minus the gradient of $(U_{S})_{\rm reg}$ with
respect to $r$ yields the force exerted by the gravitational field
on the rigid Casimir cavity. If $r >> \alpha$, we find the result
in Eq. (10) by working at the same level of approximation, so that
the vacuum contribution to the force acting on the cavity can be
interpreted as the (lack of) weight of virtual photons that are
not allowed to resonate in the cavity.
\vskip 0.3cm
\leftline {\bf 3. Towards the experimental verification}
\vskip 0.3cm
\noindent
In considering the possibility of experimental verification of the
extremely small forces linked to this effect we point out that
such measurements cannot be performed statically; this would make
it necessary to compare the weight of the assembled cavity with
the sum of the weights of its individual parts, a measure
impossible to perform. On the contrary, the measurements we are
interested in should be performed dynamically, by modulating the
force in a known way; the effect will be detected if the
modulation signal will be higher than the sensitivity of the
detector. In this spirit we focus on the sensitivity reached by
the present technology in detection of very small forces on a
macroscopic body, on earth, paying particular attention to
detectors of the extremely small forces induced by a gravitational
wave. As an example, gravitational wave signals $h$ of order
$\approx 10^{-25}$, corresponding to forces of magnitude $ \approx
5 \cdot 10^{-17} N$ at frequency of few tens of Hz, are expected
to be detected with the Virgo gravitational wave detector
presently under construction, after a month of integration time.

In the course of studying experimental possibilities of
verification of the force on a rigid Casimir cavity, we  evaluate
this force on a macroscopic body, having essentially the  same
dimensions of mirrors for gravitational wave detection and
obtained through a multi-layer sedimentation by a series of rigid
cavities. Each rigid cavity consists of two thin metallic disks,
of thickness of order 100 nm  separated by a dielectric material,
inserted to maintain the cavity sufficiently rigid. By virtue of
presently low costs and facility of sedimentation, and low
absorption in a wide range of frequencies, $SiO_{2}$ can be an
efficient dielectric material.

{}From an experimental point of view we point out that the Casimir
force has so far been tested down to a distance $a$ of about $60$
nm, corresponding to a frequency $\nu_{\rm min}$ of the
fundamental mode equal to $2.5 \cdot 10^{15} Hz$. This limit
results from the difficulty to control the distance between two
separate bodies, as in the case of measurements of the Casimir
pressure. As stated before, in our rigid case, present
technologies allow for cavities with much thinner separations
between the metallic plates, of the order of few nanometers. At
distances of order 10 nm, finite conductivity  and dielectric
absorption are expected to play an important role in decreasing
the effective Casimir pressure, with respect to the case of
perfect mirrors [5]. In this paper we discuss
experimental problems by relying on current technological
resources, considering cavities with plates' separation of 5 nm
and estimating the effect of finite conductivity by considering
the numerical results of Ref. [5]; this corresponds
to a decreasing factor $ \eta $  of about $7 \cdot 10^{-2}$ for
Al. Moreover, to increase the total force and obtain macroscopic
dimensions, $N_{l} = 10^{6}$ layers can be used, each having a
diameter of 20 cm, and thickness of 100 nm, for a total thickness
of about 10 cm. Last, one should also consider corrections
resulting from finite temperature and roughness of the surfaces,
although one might hope to minimize at least the former by working
at low temperatures.

With these figures, the total force ${\vec F}^{T}$ acting on the
body can be calculated with the help of Eq. (8), modified to take
into account the refractive index $n$ for $SiO_{2}$, the
decreasing factor $\eta$, the area A of disk-shape plates, and the
$N_{l}$ layers:
$$
{\vec F}^{T} \approx \eta N_{l} {A \pi^{2} \hbar c \over 720(n
a)^{3}} {g \over c^{2}}{\vec e}_{r} .
\eqno (13)
$$
This formula describes a static effect, while the need for a
feasible experiment makes it mandatory to modulate the force, and
various experimental possibilities  are currently under study. As
first, we are investigating the possibility of modulating $\eta$,
by varying the temperature, so as to achieve a periodic transition
from conductor to superconductor regime. In this case, making it
impossible for the cavity to have modes at dielectric absorption
frequencies, the index $n$ can be approximated to unity. By doing
so one can obtain $\eta_{\rm max}$ of order $5 \cdot 10^{-1}$
[6], and the magnitude of the force at the modulation
frequency can reach $10^{-14} \; N$. Even though such a force is
apparently more than two orders of magnitude larger than the force
which the Virgo gravitational antenna is expected to detect, we
should consider that the signal there is at reasonable high
frequency (some tens of Hz), while our calculated signal remains
at lower frequencies, i.e. tens of mHz.  Moreover, the technical
problems resulting from stress induced in changing temperature
require careful consideration before saying that modulation is
feasible. For this reason we are also estimating the possibility
of building cavities which have, by construction, some resonance
frequencies at dielectric absorption frequencies. Calculations of
efficiency of small length modulations under these conditions, to
allow for a measurable signal, are currently being performed.
\vskip 0.3cm
\leftline {\bf 4. Conclusions}
\vskip 0.3cm
\noindent
The relation of the Casimir energy with the geometry of bounding
surfaces is still under investigation,  and another open problem
of modern physics, i.e. the cosmological constant problem, results
from calculations which rely on the application of the equivalence
principle to vacuum energy [1], and this adds interest
to our calculations, that we have performed by focusing on a
Casimir apparatus.

Our original contributions are given by the evaluation of the
force acting on a macroscopic body which mimics the rigid Casimir
cavity, and by a detailed estimate of the expected order of
magnitude of such a force.

As far as we can see, there is room left for an assessment of our
investigation, i.e.: (i) how to make sure that the cavity is
sufficiently rigid; (ii) the task of verifying that corrugations
or yet other defects do not affect substantially our estimates;
(iii) how to perform signal modulation, which is still beyond our
reach.

The experimental verification of the calculated force, which was
our main concern, might be performed if the problem of signal
modulation could be solved. On the other hand, in the authors'
opinion, the order of magnitude of the calculated force can be
already of interest to demonstrate that experiments involving
effects of gravitation and vacuum fluctuations are not far from
what can be obtained with the help of present technological
resources.
\vskip 0.3cm
\leftline {\bf Acknowledgments}
\vskip 0.3cm
\noindent
The INFN financial support is gratefully acknowledged. The work of
G. Bimonte and G. Esposito has been partially supported by the Progetto
di Ricerca di Interesse Nazionale {\it SINTESI 2002}. We are indebted
to Bernard Kay and Serge Reynaud for correspondence, and to
Giuseppe Marmo for conversations. More important, we acknowledge
the lovely support of Cristina, Michela, Patrizia and Tonia.
\vskip 0.3cm
\leftline {\bf References}
\vskip 0.3cm
\noindent
\item{[1]}
R.P. Feynman, A.R. Hibbs, Quantum Mechanics and Path
Integrals, McGraw Hill, New York, 1965, pp. 244--246; for an
up-to-date presentation, with references, see P.S. Wesson,
Astrophys. J. 378 (1991) 466.

\item{[2]}
D.W. Sciama, in: The Philosophy of Vacuum, eds. S.
Saunders and H.R. Brown, Clarendon, Oxford, p. 137.

\item{[3]}
C. Moller, The Theory of Relativity, Oxford University Press,
Oxford, 1972.

\item{[4]}
L.S. Brown, G.J. Maclay, Phys. Rev. 184 (1969) 1272.

\item{[5]}
A. Lambrecht, S. Reynaud, Eur. Phys. J. D 8 (2000) 309.

\item{[6]}
E. Calloni, L. Di Fiore, G. Esposito, L. Milano, L. Rosa,
quant-ph/0109142.

\bye